# The stability and instability of the language control network: a longitudinal resting-state functional magnetic resonance imaging study

Zilong Li[a,†], Cong Liu[b,†], Xin Pan[a], Guosheng Ding[c*], Ruiming Wang[a,*]

[a] *Key Laboratory of Brain, Cognition and Education Sciences, Ministry of Education, & Guangdong Provincial Key Laboratory of Mental Health and Cognitive Science, Center for Studies of Psychological Application, School of Psychology, South China Normal University, Guangzhou, China*

[b] *Department of Psychology, Normal College & School of Teacher Education, Qingdao University, Qingdao, China*

[c] *State Key Laboratory of Cognitive Neuroscience and Learning & IDG/McGovern Institute for Brain Research, Beijing Normal University, Beijing, China*

[†] Zilong Li and Cong Liu contributed equally to this study.

[*]Address for correspondence:

| | |
|---|---|
| Ruiming Wang, PhD | Guosheng Ding, PhD |
| School of Psychology | Faculty of Psychology |
| South China Normal University | Beijing Normal University |
| 510631 Guangzhou, P. R. China. | 100875 Beijing, P. R. China. |
| E-mail: wangrm@scnu.edu.cn | E-mail: dinggsh@bnu.edu.cn |

**Abstract**

The language control network is vital among language-related networks responsible for solving the problem of multiple language switching. Researchers have expressed concerns about the instability of the language control network when exposed to external influences (e.g., Long-term second language learning). However, some studies have suggested that the language control network is stable. Therefore, whether the language control network is stable or not remains unclear. In the present study, we directly evaluated the stability and instability of the language control network using resting-state functional magnetic resonance imaging (rs-fMRI). We employed cohorts of Chinese first-year college students majoring in English who underwent second language (L2) acquisition courses at a university and those who did not. Two resting-state fMRI scans were acquired approximately 1 year apart. We found that the language control network was both moderately stable and unstable. We further investigated the morphological coexistence patterns of stability and instability within the language control network. First, we extracted connections representing stability and plasticity from the entire network. We then evaluated whether the coexistence patterns were modular (stability and instability involve different brain regions) or non-modular (stability and plasticity involve the same brain regions but have unique connectivity patterns). We found that both stability and instability coexisted in a non-modular pattern. Compared with the non-English major group, the English major group has a more non-modular coexistence pattern.. These findings provide preliminary evidence of the coexistence of stability and instability in the language control network.

# 1 Introduction

In multicultural and multilingual societies, bilinguals are required to switch between languages. This process requires bilingual language control, which refers to the cognitive abilities that minimize interference from the non-target language when bilinguals use the target language (Declerck et al., 2015). Previous studies have proposed that language control processes involve a complex neural network that includes multiple cortical and subcortical brain regions (Abutalebi & Green, 2016; Green & Abutalebi, 2013; C. Liu, de Bruin et al., 2021; R. Wang et al., 2016; Wu et al., 2019). For example, a meta-analysis of studies on bilingual language control identified eight brain regions as part of the language control network: the pre-supplementary motor area (pre-SMA), left inferior frontal gyrus (IFG), left middle temporal gyrus, left middle frontal gyrus, right precentral gyrus, right superior temporal gyrus, and bilateral caudate (Luk et al., 2012). In addition to these regions, the dorsal anterior cingulate cortex is often activated during language control and plays a role similar to that of the pre-SMA (Abutalebi et al., 2013; Abutalebi & Green, 2016; Branzi et al., 2016).

Previous studies have shown that resting-state brain connectomes are stable (Amico & Goñi, 2018; Bari et al., 2019; Chen et al., 2015; Finn et al., 2015; Kaufmann et al., 2017; J. Liu et al., 2018; Munsell et al., 2020; Ravindra et al., 2019; Yeh et al., 2016; Zhang et al., 2022; Zuo & Xing, 2014). This stability is peculiar, similar to that of a fingerprint, indicating that the brain connectome is highly variable across individuals (unique) but highly reliable (stable) at different times in each individual (Dufford et al., 2021). Notably, a high degree of stability is present in the prefrontal lobe (Finn et al., 2015; J. Liu et al., 2018; L. Liu et al., 2020; Yeh et al., 2016; Zhang et al., 2022). Furthermore, several brain regions in the language control network belong to the frontoparietal lobe. Therefore, stability may also exist in the language control network. In addition, several studies have indirectly supported the stability of speech control networks. Notably, many studies have also indicated the stability of the language control network. For example, Liu et al. (2021) found that long-term second language (L2) learning modified only the functional connectivity strength between two brain regions of the language control network, whereas the other brain regions remained unchanged. Furthermore, the extended classroom instruction did not modify the gray matter volume of the left anterior cingulate cortex or caudate, which are components of the language control network (C. Liu, Jiao, et al., 2021). Therefore, the language control network may be inferred to have a certain degree of stability; however, this has not been directly confirmed.

In addition to stability, the brain also exhibits instability. Dufford et al. (2021) revealed that the brains of 1-year-old infants are not as distinct and stable as those of adults and proposed that this instability is due to extensive and significant alterations in the infant's brain. The language control network is believed to undergo significant changes during adulthood (Antoniou, 2019). The representative theories about the instability of language control networks are the dynamic restructuring model (DRM) (Pliatsikas, 2020) and the adaptive control hypothesis (ACH) (Green & Abutalebi, 2013). DRM suggests that language learning and switching experiences lead to dynamic restructuring. The ACH suggests that the language control network adapts to the L2 learning experience (Abutalebi & Green, 2016; Green & Abutalebi, 2013). Both DRM and ACH propose that language experience shapes the language control network. This proposal is supported by the findings that the connectome of language control networks is significantly changed after long-term L2 learning (Barbeau et al., 2017; Kang et al., 2017) and that both regional brain activity and functional connectivity are sensitive to L2 experience (Zou et al. 2012; Abutalebi and Rietbergen 2014; Baum and Titone 2014; Bialystok 2014; Li et al. 2014; Tu et al. 2015; Kroll and Chiarello 2016; Pliatsikas 2020).

The language control network may be simultaneously stable and unstable, and it may seem contradictory to think that a brain network has two contrasting properties. However, Spear (2013) proposed that this is possible because there is a balance between stability and instability in the brain. Specifically, instability indicates that the brain can adapt to new experiences throughout life, whereas stability indicates that the brain can resist certain changes to maintain reliable cognition and behavioral patterns. Since stability and instability can coexist in language control networks, understanding how they coexist is essential. Fedorenko and Thompson-Schill (2014) provide valuable insights into two potential coexistence models. The first proposes that distinct functions are achieved through the combination of a core module and specific peripheral modules. Each module comprises various nodes (different brain regions). In the language control network, certain brain regions are involved in certain functions that lead to perceptual instability, whereas others are responsible for maintaining stability in language control abilities. However, the second model suggests no modularity in the language network; the entire language network performs its corresponding functions by reorganizing itself into different patterns. Brain regions within the language control network adapt to environment via a specific pattern of connectivity. This pattern must exhibit significant instability, whereas the other patterns must remain stable to fulfill certain requirements.

Therefore, we propose two alternative hypotheses regarding the coexistence of stability and instability in the language control network. The first is that the coexistence pattern is modular, with one set of nodes supporting stability and another set supporting instability (Fig. 1C). This hypothesis suggests tight inner connections/edges between nodes within the same set but loose connections between nodes from different sets. In this case, the nodes of the language control network can be modularized into two clusters, each focusing on a specific feature. Coexistence may not be modular; the entire network implements language control through a specific connection pattern, whereas no other connection pattern is involved in language control. This implies that some connections in the language control network may easily change over time, whereas others remain stable. In this case, the instability and stability of the network are embodied by separate connections/edges (Fig. 1C).

In addition, the coexistence pattern of the language control network may be modulated by L2 experience. Different levels of L2 experience can lead to different levels of brain remodeling (Caffarra et al., 2015; Cargnelutti et al., 2022; Kotz, 2009; Połczyńska & Bookheimer, 2021). The DRM also suggests that with an increase in L2 experience, instability manifests in three distinct patterns. Notably, diverse levels of linguistic proficiency elicit varying impacts on the language control network. L2 experience can influence the balance between stability and instability in the language control network. An increased bilingual experience may lead to a less stable network. However, the morphological appearance of the pattern of higher instability coexisting with lower stability has not yet been adequately investigated.

In the present study, we aimed to investigate the coexistence of stability and instability in the language control network. To achieve this, we used resting-state functional magnetic resonance imaging (fMRI) data at 1-year intervals. Further, we aimed to explore the relationship between bilingual experience and different coexistence patterns. For this, we employed datasets from native-speaking Chinese college students who majored in English and those who did not. English majors accepted long-term, diverse classroom L2 learning, whereas non-English majors received less diverse and frequent weekly classroom L2 learning. First, data from both groups were used to directly measure stability and instability in the language control network. Next, we determined whether the morphological coexistence pattern was modular. Finally, we investigated whether there were differences in the coexistence of stability and instability in the language control network of English and non-English majors.

**2 Materials and Methods**

*2.1 Participants*

Fifty-four college students were recruited for this study. The students underwent two fMRI scans for approximately 1 year. Three participants were excluded because they did not participate in the post-test session, and another three were excluded because of poor data quality owing to excessive head movements. Therefore, 49 participants were included in the final analysis (35 women; mean age=18.52 years old [standard deviation (SD)=1.17]). The 49 participants were divided into two groups: English and non-English majors. The English majors group included 20 first-year undergraduate students (18 women; mean age: 18.52 years old [SD: 0.68]) majoring in English at South China Normal University (SCNU). During the year of L2 learning, the participants accessed naturalistic, varied, and immersive L2 learning in real life, different from traditional L2 learning studies, in which specific training tasks (phonological/semantic choice tasks) under laboratory conditions were chosen (Qu et al., 2017; Wang et al., 2017). The average age at which they started learning English was 7.60 years old (SD = 3.0). This study was approved by the Research Ethics Committee of SCNU. All participants signed written informed consent forms before the experiment and received compensation for their time during the experiment.

For consistency with the categories of L1 and L2 for the English majors and the level of education in universities, 29 Chinese college students from the Southwest University Longitudinal Imaging Multimodal (SLIM), which is available online (http://fcon_1000.projects.nitrc.org/indi/retro/southwestuni_qiu_index.html) and documented in the study of Liu et al., (2017), were recruited as the non-English majors group (17 women; mean age: 19.68 years old [SD: 1.09]). They were all native Chinese speakers, and none were English majors. Compared with English majors, non-English majors received less classroom English learning in 1 year. They only accepted 1–2 h of weekly English classroom learning, mainly listening and speaking, which is far less in quantity and variety than the approximately 40 h of weekly bilingual learning for English majors. Data on the age at L2 acquisition for the non-English majors group was missing because it was unavailable in the public database. However, both English and non-English majors underwent combined Raven's test. The independent samples t-test showed no significant differences in intelligence levels between the two groups ($t$ =1.28, $df$ =48, $p$ =0.209).

*2.2 Stimuli and experimental design*

The English majors underwent two sessions of resting-state scanning within 1 year, once in the first semester of their first year (Session 1) and once in the third semester (Session 2). In both sessions, they completed a self-rating scale (range: 1–7 [1 = not proficient and 7 = very proficient]) to assess their L2 proficiency. We also employed the Oxford Placement Test (OPT) in the post-learning session to ensure the reliability of the self-rating scale. Pearson's correlation analyses revealed a significant positive correlation between the self-rated English proficiency and OPT scores ($r = 0.355$, $p < 0.05$), confirming that the self-rating scale was somewhat reliable. The non-English majors group also underwent two sessions of scanning within 1 year. Due to the lack of assessment of L2 proficiency in the SLIM, non-English majors had no English proficiency scores. The close rankings of the two universities in the U.S. News & World Report Best Global Universities Rankings (Southwest University ranked 667, whereas SCNU ranked 770) and ShanghaiRanking's Best Chinese Universities Rankings (Southwest University ranked 72, whereas SCNU ranked 79) suggest that student levels are relatively similar. Therefore, it is unlikely that the non-English majors group at one university will have a higher proficiency level than the English majors group at the other university. In addition, English majors are believed to have greater L2 proficiency than non-English majors because they receive approximately 20 times longer and richer L2 education.

*2.3 Imaging data acquisition*

Resting-state fMRI scanning lasted for 8 min. During scanning, participants were asked to close their eyes, rest, think of nothing, and remain still. MRI images were acquired using a 3T Siemens Trio scanner with a 12-channel phase-array head coil at the SCNU. Functional images were acquired using a T2*-weighted gradient-echo echo planar imaging (EPI) sequence with the following sequence parameters: repetition time (TR) = 2000 ms, time to echo (TE) = 30 ms, flip angle = 90°, field of view (FOV) = 204 × 204 mm$^2$, matrix = 64 × 64, slice thickness = 3.5 mm, interslice gap = 0.5 mm, and voxel size = 3 × 3 × 3.5 mm$^3$. High-resolution brain structural images were acquired for all participants using a three-dimensional T1-weighted MP-RAGE sequence (TR = 1900 ms, TE = 2.52 ms, flip angle = 9°, FOV = 256 × 256 mm$^2$, matrix = 204 × 204, slice thickness = 1 mm, and voxel size = 1 × 1 × 1 mm$^3$).

Images of the non-English majors group were acquired using a 3T Siemens Trio scanner with a 12-channel phase-array head coil at Southwest University. Functional images were obtained using a T2*-weighted gradient-echo EPI sequence (TR = 2000 ms, TE = 30 ms, flip angle = 90°, FOV = 220 × 220 mm$^2$, matrix = 64 × 64, slice thickness = 3.0 mm, interslice gap = 0.5 mm, and voxel size = 3.4 × 3.4 × 3.0 mm$^3$).

*2.4 Imaging data preprocessing*

Data were preprocessed using the GRETNA toolbox based on SPM 12 ([www.fil.ion.ucl.ac.uk/spm/software/spm12/](www.fil.ion.ucl.ac.uk/spm/software/spm12/)) (Wang et al., 2015). The preprocessing steps included (1) removing the first 10 images, (2) slice timing correction, (3) realignment, (4) normalization to MNI space, (5) resampling to 3 mm isotropic voxels, (6) spatial smoothing with a 6 mm Full Width at Half Maximum Gaussian kernel, (7) removal of linear drift, (8) regressing out 24-parameter head motion profiles (Friston et al., 1996) and the global, white matter, and cerebrospinal fluid signals, and (9) temporal filtering with frequency of 0.008–0.083 Hz.

*2.5 Network construction*

Based on previous studies and meta-analyses (Green & Abutalebi, 2013; Luk et al., 2012; Sun et al., 2019), we selected 11 regions of interest (ROIs) as nodes to construct a language control network ( Fig. 1A). All nodes were defined as spheres with a radius of 6 mm. The mask for the ROIs is based on the brain mask provided by GRETNA (Wang et al., 2015) toolbox. The coordinates of the nodes in the networks are presented in Table 1.

Table 1. Node coordinates of language control network in MNI space.

| No | Language control network | |
|---|---|---|
| | Node | Coordinate |
| 1 | L_MFG (BA46) | (-48, 49, 27) |
| 2 | Pre_SMA | (2, 4, 64) |
| 3 | L_IFG(BA47) | (-33, 25, -13) |
| 4 | R_PrCG | (46, -5, -29) |
| 5 | R_Caudate | (17, 10, 11) |
| 6 | L_MTG | (-54, -43, -11) |
| 7 | L_IFG (BA44) | (-52, 21, 4) |
| 8 | R_STG | (55, -19, -8) |
| 9 | L_MFG (BA9) | (-44, 8, 31) |
| 10 | L_Caudate | (-11, 22, -7) |
| 11 | dACC | (0, 6, 44) |

Functional connectivity matrices were analyzed using the GRETNA (Wang et al., 2015)

toolbox. The steps of functional connection (FC) network construction are as follows: (1) the BOLD time series of all voxels within each node was averaged; (2) the inter-nodal Pearson's correlation of BOLD time series was computed; (3) Fisher r-to-Z transformation.

*2.6 The tests of stability and instability*

Functional connectome fingerprint identification (fingerprinting) was performed to determine whether the language control network was stable. In addition, instability was evaluated using the classifier (Fig. 1B).

*2.6.1 Functional connectome fingerprint identification*

The stability of the entire language control network was evaluated using functional connectome fingerprint identification (Finn et al., 2015). This is also known as "fingerprinting" and can individually identify different people based on their functional connections (Bari et al., 2019; Dufford et al., 2021; Finn et al., 2015; Ravindra et al., 2019).

Fingerprinting was implemented through the following steps: (1) selecting a participant and computing the Pearson correlation to connectivity networks between this participant in the pre-session and all other participants in the post-session. If two networks from the same participant have the highest correlation coefficient, this participant's identification is successful; (2) calculating the accuracy of everyone and then adding them up to get a total accuracy (the accuracy of pre-session-2); (3) computing another total accuracy after switching the pre- and post-session (the accuracy of post-session-1), and then computing the average value between the accuracy of pre-session-2 and post-session-1 as the fingerprinting accuracy to ensure the brevity of results; (4) permutation testing: randomizing the index of the subject in one set and computing the total accuracy. This permutation test was repeated 1000 times. Statistical significance was defined as the percentage of the accuracy distribution higher than the original accuracy.

*2.6.2 Network classifier*

The classifier is a kind of algorithm in machine learning and can automatically categorize data into one or more sets of "classes." In the present study, the classifier was used to detect non-linear group differences in the language control network between pre- and post-session and categorized the resting-state fMRI data into pre- and post-session classes.

The pre-session network was labeled -1, whereas the post-session network was labeled 1. Because of the small number of participants, leave-one-out cross-validation was used to evaluate the generalizability of the model (Tang et al., 2018). First, a participant's pre- or post-test data were employed as a test set, and then the model was trained using the remaining data. After obtaining the model, we tested whether it could correctly predict whether the test set belonged to the pre- or post-test. Analogously, each participant's pre- and post-test data were used as a test set to verify whether the model could successfully predict the pre- and post-tests. The final success percentage was used as classification accuracy. The support vector classifier fits the training set, and the model is cross-validated on the test set. Our classifier was based on the LIBSVM (Chang & Lin, 2011) toolbox, with a linear kernel and otherwise default parameters.

The permutation was repeated 1000 times to obtain a p-value for accuracy. In the present study, we selected the classification accuracy of a Support Vector Machine (SVM) as an indicator of instability. This is because SVMs fulfill the requirement of "the opposite of stability." Specifically, high stability, as measured using fingerprinting, implies significant interindividual variability, which remains stable over time. In contrast, instability implies small individual differences that vary significantly over time. The method of classifying SVM is to construct a hyperplane, separating the participants at different time points in two places. The further the participants at different time points are from the hyperplane, the higher the classification accuracy rate of SVM, whereas the smaller the individual differences within each of the different time points, the less likely the data selected as the test set to be misclassified. Therefore, when there is a difference between the pre- and post-tests and the individual differences within the pre- and post-tests are small, SVM has the best classification effect and the highest accuracy rate. In addition, because SVM maps the data to a high-dimensional space, it can detect non-linear differences at different time points and more accurately reflect differences at different time points.

*2.7 The coexisting pattern analyses*

A series of analyses were performed to detect the morphological coexistence pattern of stability and instability in the language control network. First, we created a filter based on Differential Power (DP) to separate the edges of the language control network into two parts: stability and instability (Fig. 1C). Next, we compared the filter results with computer simulations to determine whether the morphological coexistence pattern was modular (Figure 1D).

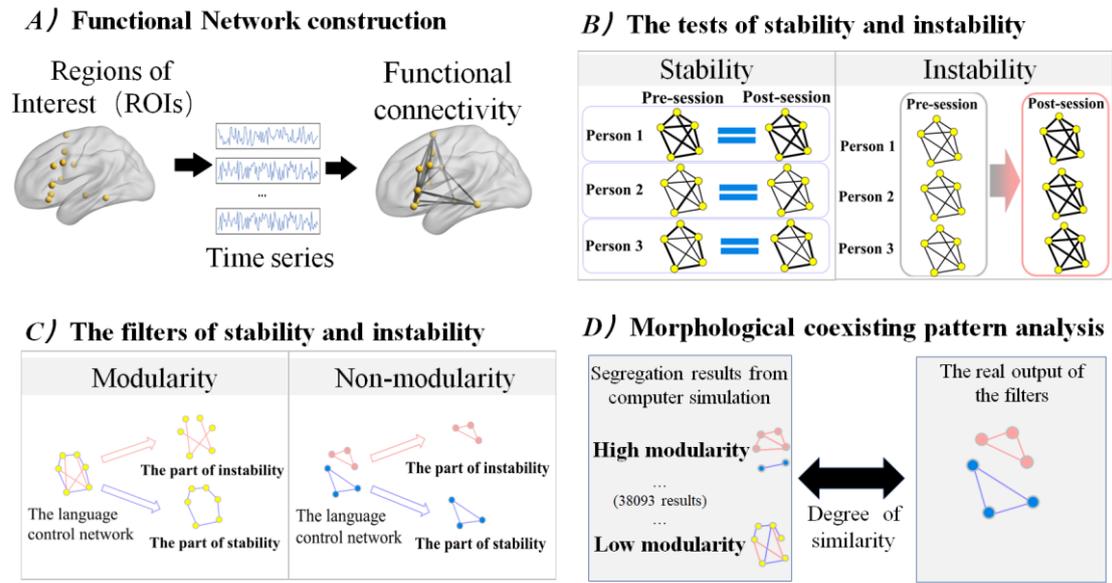

**Figure 1.** The pipeline of the morphological coexistence pattern analysis. (A) Time series of brain activation in regions of interest (ROIs) were extracted, and then functional connectivity matrices were calculated. (B) The tests of stability and instability. The connectome fingerprinting evaluated stability, measuring the extent to which individual differences in brain connectomes of participants do not differ between sessions. The classifier evaluated instability, calculating the difference between the two sessions. (C) The parts of stability and instability were separated from the entire language control network using the filters. There are two possibilities regarding the segregation: modularity and non-modularity. (D) The method for quantifying the modularity of coexistence. The computer simulated various kinds of coexistence patterns and then calculated the modularity index. The degree of similarity between actual and simulated coexistence patterns was determined. The correlation between the degree of similarity and modularity index quantified the extent of modularity.

*2.7.1 The stability and instability filters*

The filter is responsible for filtering out the set of connections that best characterizes stability and instability in the language control network, as shown in Fig. 1C. Finn et al. (2015) reported using DP to determine the uniqueness and stability of certain edges. Edges with a high DP exhibited high fingerprinting accuracy. The set of these edges was defined as the stable part of the language control network (Figure 1C). According to the DP algorithm, it can be further inferred that the edges with a low DP are low in uniqueness and unstable or change more obviously. This implies that they exhibit high instability.

Furthermore, classifier accuracy can be used to measure the degree of instability. Consequently, the set of edges with a low DP, which had the highest classifier accuracy, were defined as the unstable part of the language control network (Figure 1C). Therefore, we built a three-part filter to determine the stable and unstable parts.

The first step is the DP analysis, which produces a DP matrix to indicate subsequent steps. The second is the stability filter, which is based on a DP matrix and fingerprinting. The third step is the instability filter, which is based on the DP matrix and classifier.

*2.7.1.1 Differential power analysis*

Finn et al. (2015) indicated DP as an indicator to quantify the uniqueness of each edge within a network. A high DP indicates that the edge in one set (pre-session) is similar to the same edge in another set (post-session) but different from the other edges in the post-session. The DP analysis was implemented using custom code. We have largely maintained the algorithm of Finn et al. (2015) but adapted it slightly according to the data features in the present study to prevent infinite DP values. The steps for the DP analysis are as follows.

(1) Computing consistency of edges

$X^{R1}$ and $X^{R2}$ indicate two normalized functional network matrices from pre- and post-session, respectively. $i$ indicates subject, $j$ indicates subject different from $i$, $e$ indicates edge, and $M$ is the total number of edges in one network. $\varphi_{ii}$ indicates a vector of correspondence when the subject subscript is matched, whereas $\varphi_{ij}$ and $\varphi_{ji}$ indicate vectors of correspondence when subscript is unmatched. The specific formula is as follows:

$$\varphi_{ii}(e) = X_i^{R_1}(e) * X_i^{R_2}(e), e = 1,2,3,\cdots,M$$

$$\varphi_{ij}(e) = X_i^{R_1}(e) * X_j^{R_2}(e), e = 1,2,3,\cdots,M, i \neq j$$

$$\varphi_{ji}(e) = X_j^{R_1}(e) * X_i^{R_2}(e), e = 1,2,3,\cdots,M, i \neq j$$

*(2) Computing empirical probability*

If $\varphi_{ii}(e)$ is equal to $\varphi_{ij}(e)$ or $\varphi_{ji}(e)$, edge $e$ does not help to distinguish an individual from others. Therefore, an edge contributes to fingerprinting if it satisfies the following conditions only:

$$\varphi_{ii}(e) > \varphi_{ij}(e) \text{, and } \varphi_{ii}(e) > \varphi_{ji}(e) \text{ , } i \neq j$$

Based on the above conditions, we use an empirical probability $P_i(e)$ to qualify the DP of identification:

$$P_i(e) = \frac{(|\varphi_{ii}(e) < \varphi_{ij}(e)| + |\varphi_{ii}(e) < \varphi_{ji}(e)|)}{2(n-1)}$$

Where n denotes the number of participants. $P_i(e)$ is considered the same as the p-value in a standard statistical test. The lower the $P_i(e)$, the better the edge $e$ distinguishes the subject $i$. Because the next step is logarithmic conversion, if $P_i(e)$ of a certain subject is zero, the outcome of the next step will be positive infinite. When $n$ is relatively small, a positive infinity appears easily and covers the DP of the others. Extreme value makes subsequent analysis meaningless; therefore, we set a minimum of $P_i(e)$ [min $P_i(e)$] to avoid positive infinity. Setting the minimum is a common method. To ensure min $P_i(e)$ does not interfere with the analysis, its specific value must stratify the following conditions:

$$0 < \min P_i(e) < \frac{1}{2(n-1)}$$

In the present study, we choose 0.01 as min $P_i(e)$.

*(2) Logarithmic conversion*

The total DP of a certain edge across all subjects is defined by the DP measure as follows:

$$DP(e) = \sum_i \{-\ln(P_i(e))\}$$

Finally, a DP matrix of the edges was constructed. The larger the DP of a certain edge, the higher the inter-subject difference but the lower the inter-session difference. This tendency is supported by the results of the subsequent analysis.

*2.7.1.2 The instability filter*

The instability filter was implemented using custom code based on the partial functions of SPM 12 and LIVSVM (Chang & Lin, 2011) toolbox. The steps for this filter are as follows.

(1) Constructing network masks with different sparsities.

Given the DP matrix obtained from the original network, the edges with the first $m\%$

DP were removed. However, no meaningful conclusions could be drawn from the isolated connections. If one node cannot reach the third node through two edges, the filtered edges cannot form a network to transport information. Therefore, we ran 10,000 Monte Carlo simulations and found that the probability of this occurrence was zero when the number of edges exceeded six. Therefore, the m% was 6/55 (10.9%). Consequently, we retained 10% of the edges to ensure that they formed a complete network rather than being isolated. Therefore, $m$ is a vector with values ranging from 10 to 90 (with a step size of 1). The reason for not starting at 0 is that m needs to be aligned with n in the mask build of the stablility filter. The mask construction in the stablility filter also need to ensure that edges form a loop, meaning that the stablity filter needs to keep at least the top 10% of edges with the highest DP values. Subsequently, the remaining edges were used to construct 91 masks.

(2) Classifier.

Networks with different sparsities were constructed using masks. The classification accuracy and p-value for each sparsity were calculated. The parameters and steps for the classifier were the same as those previously described. The network with the smallest p-value and highest classification accuracy was considered unstable.

*2.7.1.3 The stability filter*

The stability filter was employed using custom codes based on the partial functions of SPM 12 and previously described codes (Finn et al., 2015).

The steps for this filter are somewhat similar to those for the instability filter and are as follows.

(1) Constructing network masks of different sparsities.

Unlike the instability filter, the edges with the first $n$% DP were retained. $n$ is a vector with values ranging from 10 to 90 (with a step size of 1).

(2) Fingerprinting.

Networks with different sparsities were constructed using the masks from Step 1. Connectome fingerprinting of different sparsities was performed, and p-values were obtained from the permutation test. In this filter, the first network with the highest fingerprinting accuracy was considered stable.

*2.7.2 Morphological coexistence pattern analysis*

The outputs of the previous filter can provide an intuitive visualization of the coexistence patterns of stability and instability; however, statistical evidence that the coexistence pattern is modular remains lacking. Therefore, we used a method based on the Fast Newman algorithm to detect the morphological coexistence patterns.

The analysis consisted of the following steps. The first step was to reveal all possible segregations in which the language control network could be divided into two modules. We accessed every way through which the 11 brain regions could be divided into two modules. We ensured that the number of nodes in each module was greater than two. The second step was matrix calculation. We calculated the average functional connectivity matrix of all the participants to obtain a matrix representing the actual matrix. According to the definition of modularity, 38093 kinds of artificial matrices were created for all the simulated segregations. For artificial matrices, nodes are connected within the same module but not between different modules. The third step was indicator calculation. The fast Newman algorithm was used to calculate the degree of modularity (Q-value) for all segregations. We then calculated the Jensen-Shannon (JS) divergence between every artificial and actual matrices. Next, we used the permutation test to obtain the P-values for each artificial matrix (Guilbeault et al., 2020; Menéndez et al., 1997). A lower JS divergence indicates greater similarity between the two matrices. Therefore, the two patterns of the artificial and actual matrices were the same. The P-value of the permutation test reflects the probability that the JS value is higher than the random level. The higher the P-value, the greater the similarity between the actual and artificial matrices. Finally, we calculated the Pearson correlation coefficient between the Q-value and P-value. The higher the correlation coefficient, the greater the similarity between the actual and highly modular artificial matrices. Therefore, a higher correlation coefficient suggests that the coexistence pattern is more modular. The fMRI data from English and non-English majors underwent the above calculations separately. The Cocor toolbox (Diedenhofen & Musch, 2015) was used to compare the two correlation coefficients and test whether there was a difference between English and non-English majors.

**3 Results**

*3.1 Results for the test of stability and instability*

For English majors, the fingerprinting accuracy of the language control network was 40% ($p < 0.001$), and the classifier accuracy was 8/20 (40%, $p = 0.88$). For non-English majors, the fingerprinting accuracy was 50% ($p < 0.001$), and the classifier accuracy was 8/29 (27.59%, $p$ = 0.99). Neither English nor non-English majors had a higher classifier accuracy than the random level (50%). Furthermore, English and non-English majors exhibited stability above the random level, whereas instability was not above the random level (Fig. 2A).

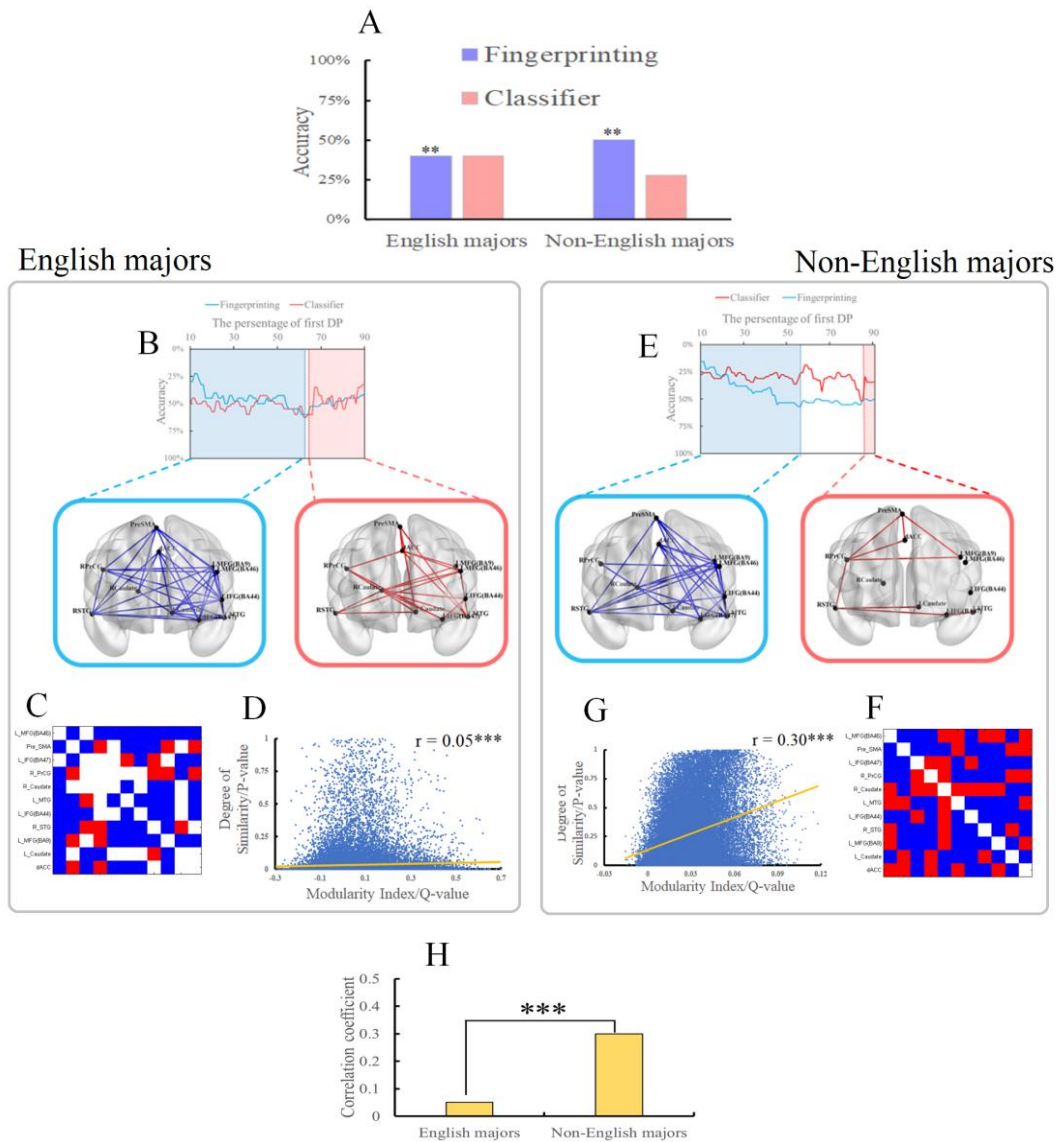

**Figure 2.** The results of morphological coexistence pattern analysis. (A) The results of tests of stability and instability in the entire language control network. (B) The upper

part is the filter for English majors. The blue area is the stable part, whereas the red area is the unstable part. The lower part is the edges of the stable and unstable parts in English majors displayed from a coronal view of the brain. (C) The matrix complex of the stable and unstable parts of English majors. The blue blocks make up the stable part, whereas the red blocks make up the unstable part. (D) The correlation between the modularity index (Q-value) and similarity index (P-value) in English majors. (E) The upper part is the filter for non-English majors. The blue area is the stable part, whereas the red is the unstable part. The lower part is the edges of the stable and unstable parts from non-English majors displayed from a coronal view of the brain. (F) The matrix complex of stable and unstable parts of non-English majors. (G) The correlation between the modularity index (Q-value) and similarity index (P-value) in non-English majors. (H) The comparison of correlation coefficients between English and non-English majors. **$p < 0.01$, ***$p < 0.001$.

*3.2 Results of the morphological coexistence pattern analysis*

For English majors, the set of edges with the highest classifier accuracy (62.5%) and smallest p-value (0.134) was defined as the unstable part. This consisted of connections with the lowest DPs (red area in Fig. 2B). The set of edges with the highest fingerprinting accuracy (60%, $p < 0.001$) was defined as the stable part. This consisted of connections with the highest DPs (blue area in Fig. 2B).

For non-English majors, the sets of edges with the highest classification accuracy (51.7%, $p = 0.42$) and fingerprinting accuracy (50%, $p < 0.001$) were defined as the unstable and stable parts, respectively, which consisted of connections with the lowest and highest DPs, respectively (red and blue areas, respectively, in Fig. 2E).

The stable and unstable parts comprise different connections in English majors. The unstable part had 20 edges (red links from B in Fig. 2 and red blocks from C in Fig. 2), accounting for 36.4% of all the edges in the language control network. The stable part had 35 edges (blue links from B in Fig. 2 and blue blocks from C in Fig. 2), accounting for 63.6% of all the edges. Notably, both parts involve all the nodes. Finally, the correlation coefficient between Q- and P-values was 0.05 (Fig. 2D). This low correlation coefficient indicated that the actual coexistence pattern was not similar to the highly modular simulated coexistence patterns. Therefore, the coexistence pattern for English majors did not exhibit high modularity.

For non-English majors, the unstable part had nine edges and did not cover all the nodes (red links from E in Fig. 2 and red blocks from F in Fig. 2), accounting for 16.4% of the language control network. The inter-session network had 31 edges (blue links from E in Fig. 2 and blue blocks from F in Fig. 2), accounting for 56.4% of the language control network. There were no isolated nodes in the networks, and only 72.7% of the edges were involved in the construction of the two subnetworks. The correlation coefficient between Q- and P-values was 0.3 (Fig. 2G). The correlation coefficient for non-English majors was significantly higher than for English majors ($Z = 35.35$, p-value <0.001. This indicates that the coexistence pattern in non-English majors still has low modularity but is more modular than that in English majors (Fig. 2H).

These results indicated that the coexistence patterns of stability and instability were not modular. Furthermore, the coexistence patterns of English and non-English majors were morphologically different. Images A, B, C, and D were drawn using BrainNet Viewer (Xia et al., 2013).

## 4 Discussion

We explored the coexistence patterns of stability and instability in the language control network. We found that the resting-state language control network exhibited moderate stability, below the 80-95% identification accuracy of the adult whole brain. In addition, English majors had less stability than non-English majors. The language control network was stable or unstable in a non-modular pattern, indicating that one part of the network did not exhibit stability, whereas the other exhibited instability. Notably, the coexistence pattern in English majors exhibited a greater degree of non-modularity, with all brain regions involved in both the stable and unstable parts of the network. In contrast, non-English majors only had a stable part involving all brain regions.

*4.1 Stability and instability in the language control network*

There is a lack of direct research on whether the language control network has the same high degree of stability and uniqueness as the whole brain. Instability has been discussed in several studies; however, there is a lack of direct quantitative research, especially when stability and instability are examined in the same framework. From an overall network perspective, the language control network exhibits a moderate degree of fingerprinting accuracy. The stability of the whole-brain network has been reported to be 80-95% in previous studies (Finn et al., 2015; J. Liu et al., 2018, 2018; Ravindra et al., 2019). The fingerprinting accuracy of 1-year-old infants was reported to be 26.6% (Dufford et al., 2021). The accuracy of fingerprinting within the language control network is approximately 40%–50%. This indicates that the language control network neither displays the same level of stability as that in the adult brain nor exhibits widespread changes as in the infant brain. Instead, it appears to be in a fragile state of equilibrium, neither fully solidified nor completely fluid.

In the present study, the classification accuracy was 40% for English majors compared with 27.59% for non-English majors; however, neither surpassed the randomized level. This indicates that the language control network did not undergo significant changes before and after 1 year. Consequently, the language control network was moderately stable and not particularly unstable overall. However, the instability of language control networks and the influence of L2 learning have been supported by several studies (Abutalebi & Rietbergen, 2014; Li et al., 2014; C. Liu, de Bruin, et al., 2021; Mei et al., 2015; Pliatsikas, 2020). These are not conflicting results because we measured instability at the scale of the entire language control network, whereas previous studies mostly found significant instability at a scale smaller than the entire network (at the scale of a single brain region or connection). The fact that few connections or brain regions in the language control network underwent significant changes does not necessarily indicate significant instability throughout the entire language control network. Therefore, moderate stability and instability in the language control network indicate that changes have occurred but are not significant, widespread, or complete. Therefore, some connections or brain regions in the language control network may have changed, but not in the majority, whereas others remained stable. This reflects the idea that, compared to the typically stable adult brain and unstable infant brain, the language control network has a balance of stability and instability, indicating that both are in an intermediate state with no clear advantage. Understanding what this equilibrium entails and whether stability and instability occupy a portion of the brain area respectively or whether all brain areas are involved in stability and instability but with different connectivity patterns is essential. The morphological coexistence pattern analysis revealed these insights. In addition, English majors exhibited lower levels of stability and higher levels of instability than non-English majors. This could be a result of the fact that English majors are expected to exercise greater language control, which leads to more changes in brain regions and connections. This is supported by several studies (Abutalebi & Rietbergen, 2014; Barbeau et al., 2017; Bialystok, 2014, 2021; Kroll & Chiarello, 2016, 2016; Li et al., 2014; C. Liu, de Bruin, et al., 2021; C. Liu, Jiao, et al., 2021; Pliatsikas, 2020; Tu et al., 2015; Zou et al., 2012).

*4.1.1 Morphological coexistence pattern*

The filters separated the stable and unstable parts from the entire language control network and then detected whether the coexistence pattern was modular. If the actual coexistence pattern is modular, it will be more similar to highly modular artificial patterns than to those with low modularity. In the present study, we plotted the Q-value (the modularity of certain artificial patterns) on the x-axis and the P-value (the similarity between the actual pattern and certain artificial patterns) on the y-axis. The highly modular artificial pattern is located further to the right of the horizontal axis. If the highly modular artificial pattern is on the right side of the vertical axis, the farther up the vertical axis, the greater the similarity between the artificial and actual patterns. Consequently, the actual pattern is modular when the Q- and P-values show a significantly large positive correlation. In the present study, the stable and unstable parts did not occupy a portion of each node to form two modules. Both parts were constructed by carving different connections. The correlation coefficients between Q- and P-values were low for English and non-English majors. These results indicate that the coexistence patterns of stability and instability are non-modular.

In addition, the coexistence pattern was less modular in English majors than in non-English majors. Because the data used in the present study were obtained from a publicly available database, the participants were not closely matched. Therefore, it is inappropriate to assume that the observed difference is solely due to disparities in classroom L2 learning. However, we controlled for first language proficiency, school rank, and fluid intelligence in the English majors group compared with the non-English majors group. We believe that the difference in their abilities may be attributed to the considerably richer and more diverse English education that English majors receive compared with non-English majors. The L2 learning of non-English majors is less in quantity and quality. This is similar to the initial exposure stage in the DRM. The DRM suggested that a specific part of the node was first affected by L2 experience during the initial exposure stage. The present study found that instability in non-English majors did not involve all language control network nodes. In contrast, instability in English majors involved all nodes rather than part of the nodes. This may indicate that the influence of L2 experience is not limited to certain brain regions but to all nodes of the language control network. Synthesizing the coexistence patterns in the language control networks of English and non-English majors, we propose a hypothesis regarding the impact of classroom L2 learning on the language control network. The mechanism through which L2 learning affects the language control network may be similar to the gradual construction of multiple expressways. The brain regions are similar to many cities. For non-English majors, the year they received less L2 learning may be similar to the initial stages of building a highway network. During the initial stages of expressway construction, only a few cities are connected by expressways. For English majors, the situation is similar to the later stages of an expressway network construction project. In the later stages of expressway construction, specific highways are connected to all cities. However, not every city is directly connected; some cities are only indirectly connected to other cities because of construction costs or other reasons. Therefore, we specified our hypotheses on how L2 learning impacts the language control network. Initially, only a few connections of specific brain regions were affected by L2 learning. However, some connections in each brain region were eventually affected. Therefore, the stable and unstable parts shared some nodes in non-English majors, whereas they shared all nodes in English majors. If the nodes are considered chemical elements, and the connections are considered bonds, the stable and unstable parts are similar to a pair of isomers in chemistry (compounds with identical chemical formulae but different structures) (Regalado et al., 2013). This result is inconsistent with that of the DRM. In the DRM, only a few specific brain regions change at each stage. This disparity may be due to the complex mapping of changes in brain function and structure. The DRM focused on the dynamic restructuring of gray and white matter, whereas we focused on the functional connectome. Complex mapping between brain structure and functional connectivity may be responsible for this inconsistency.

*4.2 The temporal complexity of the language control network*

Tononi et al. (1994) proposed that a brain network is complex when neither segregated nor integrated. The more balanced the two features are, the higher the complexity of the network as a whole. Notably, many studies have adopted this definition (Bassett & Gazzaniga, 2011; Marshall et al., 2016). Furthermore, many phenomena related to brain network complexity have been identified, such as the small world (Sporns & Honey, 2006), modularity (Gallos et al., 2012; Liang et al., 2021), and rich club (van den Heuvel & Sporns, 2011). All these phenomena reflect the complexity from a morphological or spatial perspective by showing how integration and separation coexist in brain networks. Integration and separation, as well as stability and instability, are at the two ends of a continuum. Therefore, we can preliminarily propose "temporal complexity" using analogy with morphological or spatial complexity. Specifically, the brain network is a temporal complex that is neither stable nor unstable. In the present study, the language control network may be considered to have high temporal complexity. The entire language control network exhibited neither high stability nor instability. In the coexistence pattern, these two features dominate many connections, resulting in two sets of connections involving many brain regions. Specific temporal complexity phenomena, such as two parts sharing all the nodes, were also found. However, the present study is only a preliminary exploration and does not provide very strong evidence. The temporal complexity of language control and other functionally specific brain networks remains to be further explored.

*4.3 Limitations*

First, the present study had a small sample size. Previous studies on resting-state functional connectivity had a large sample size; however, only 49 participants were included in this study. This may have weakened the generalizability and robustness of our results and conclusions. Second, learning for over 1 year was affected by many irrelevant variables. Randomization and other methods were used to eliminate interference; however, there is no guarantee that classroom L2 learning was the only independent factor.

## 5 Conclusions

This study explored the coexistence of stability and instability in the language control network. The language control network exhibited moderate stability and instability. In the language control network, stability and instability coexist in a non-modular manner. Notably, English majors exhibited more significant non-modularity than non-English majors.

*Data and code availability statement*

The dataset analyzed in the present study and the code of morphological coupled model analysis are available from the corresponding authors through the Harvard Dataverse

(https://dataverse.harvard.edu/dataset.xhtml?persistentId=doi:10.7910/DVN/ONRHXA).

*Declaration of Competing Interest*

The authors have no competing interests to declare.

*Funding*

This study was funded by the National Natual Sciences Foundation of China (32371114).

*Acknowledgement*

The author would like to thank Editage for technical editing of the manuscript.

**References**

Abutalebi, J., Della Rosa, P. A., Ding, G., Weekes, B., Costa, A., & Green, D. W. (2013). Language proficiency modulates the engagement of cognitive control areas in multilinguals. *Cortex*, *49*(3), 905–911. https://doi.org/10.1016/j.cortex.2012.08.018

Abutalebi, J., & Green, D. W. (2016). Neuroimaging of language control in bilinguals: Neural adaptation and reserve. *Bilingualism: Language and Cognition*, *19*(4), 689–698. https://doi.org/10.1017/S1366728916000225

Abutalebi, J., & Rietbergen, M. J. (2014). Neuroplasticity of the bilingual brain:

Cognitive control and reserve. *Applied Psycholinguistics*, *35*(5), 895–899. https://doi.org/10.1017/S0142716414000186

Amico, E., & Goñi, J. (2018). The quest for identifiability in human functional connectomes. *Scientific Reports*, *8*(1), 8254. https://doi.org/10.1038/s41598-018-25089-1

Antoniou, M. (2019). The Advantages of Bilingualism Debate. *Annual Review of Linguistics*, *5*(1), 395–415. https://doi.org/10.1146/annurev-linguistics-011718-011820

Barbeau, E. B., Chai, X. J., Chen, J.-K., Soles, J., Berken, J., Baum, S., Watkins, K. E., & Klein, D. (2017). The role of the left inferior parietal lobule in second language learning: An intensive language training fMRI study. *Neuropsychologia*, *98*, 169–176. https://doi.org/10.1016/j.neuropsychologia.2016.10.003

Bari, S., Amico, E., Vike, N., Talavage, T. M., & Goñi, J. (2019). Uncovering multi-site identifiability based on resting-state functional connectomes. *NeuroImage*, *202*, 115967. https://doi.org/10.1016/j.neuroimage.2019.06.045

Bassett, D. S., & Gazzaniga, M. S. (2011). Understanding complexity in the human brain. *Trends in Cognitive Sciences*, *15*(5), 200–209. https://doi.org/10.1016/j.tics.2011.03.006

Branzi, F. M., Calabria, M., Boscarino, M. L., & Costa, A. (2016). On the overlap between bilingual language control and domain-general executive control. *Acta Psychologica*, *166*, 21–30. https://doi.org/10.1016/j.actpsy.2016.03.001

Caffarra, S., Molinaro, N., Davidson, D., & Carreiras, M. (2015). Second language syntactic processing revealed through event-related potentials: An empirical review. *Neuroscience & Biobehavioral Reviews*, *51*, 31–47. https://doi.org/10.1016/j.neubiorev.2015.01.010

Cargnelutti, E., Tomasino, B., & Fabbro, F. (2022). Effects of Linguistic Distance on Second Language Brain Activations in Bilinguals: An Exploratory Coordinate-Based Meta-Analysis. *Frontiers in Human Neuroscience*, *15*, 744489. https://doi.org/10.3389/fnhum.2021.744489

Chang, C.-C., & Lin, C.-J. (2011). LIBSVM: A library for support vector machines. *ACM Transactions on Intelligent Systems and Technology*, *2*(3), 1–27. https://doi.org/10.1145/1961189.1961199

Chen, B., Xu, T., Zhou, C., Wang, L., Yang, N., Wang, Z., Dong, H.-M., Yang, Z., Zang, Y.-F., Zuo, X.-N., & Weng, X.-C. (2015). Individual Variability and Test-Retest Reliability Revealed by Ten Repeated Resting-State Brain Scans over One Month. *PLOS ONE*, *10*(12), e0144963. https://doi.org/10.1371/journal.pone.0144963

Declerck, M., Koch, I., & Philipp, A. M. (2015). The minimum requirements of language control: Evidence from sequential predictability effects in language switching. *Journal of Experimental Psychology: Learning, Memory, and Cognition*, *41*(2), 377–394. https://doi.org/10.1037/xlm0000021

Diedenhofen, B., & Musch, J. (2015). cocor: A Comprehensive Solution for the Statistical Comparison of Correlations. *PLoS ONE*, *10*, e0121945.

https://doi.org/10.1371/journal.pone.0121945

Dufford, A. J., Noble, S., Gao, S., & Scheinost, D. (2021). The instability of functional connectomes across the first year of life. *Developmental Cognitive Neuroscience*, *51*, 101007. https://doi.org/10.1016/j.dcn.2021.101007

Fedorenko, E., & Thompson-Schill, S. L. (2014). Reworking the language network. *Trends in Cognitive Sciences*, *18*(3), 120–126. https://doi.org/10.1016/j.tics.2013.12.006

Finn, E. S., Shen, X., Scheinost, D., Rosenberg, M. D., Huang, J., Chun, M. M., Papademetris, X., & Constable, R. T. (2015). Functional connectome fingerprinting: Identifying individuals using patterns of brain connectivity. *Nature Neuroscience*, *18*(11), 1664–1671. https://doi.org/10.1038/nn.4135

Friston, K. J., Williams, S., Howard, R., Frackowiak, R. S. J., & Turner, R. (1996). Movement-Related effects in fMRI time-series: Movement Artifacts in fMRI. *Magnetic Resonance in Medicine*, *35*(3), 346–355. https://doi.org/10.1002/mrm.1910350312

Gallos, L. K., Makse, H. A., & Sigman, M. (2012). A small world of weak ties provides optimal global integration of self-similar modules in functional brain networks. *Proceedings of the National Academy of Sciences*, *109*(8), 2825–2830. https://doi.org/10.1073/pnas.1106612109

Green, D. W., & Abutalebi, J. (2013). Language control in bilinguals: The adaptive control hypothesis. *Journal of Cognitive Psychology*, *25*(5), 515–530. https://doi.org/10.1080/20445911.2013.796377

Guilbeault, D., Nadler, E. O., Chu, M., Lo Sardo, D. R., Kar, A. A., & Desikan, B. S. (2020). Color associations in abstract semantic domains. *Cognition*, *201*, 104306. https://doi.org/10.1016/j.cognition.2020.104306

Kang, C., Fu, Y., Wu, J., Ma, F., Lu, C., & Guo, T. (2017). Short-term language switching training tunes the neural correlates of cognitive control in bilingual language production: Language Training Modulates Cognitive Control. *Human Brain Mapping*, *38*(12), 5859–5870. https://doi.org/10.1002/hbm.23765

Kaufmann, T., Alnæs, D., Doan, N. T., Brandt, C. L., Andreassen, O. A., & Westlye, L. T. (2017). Delayed stabilization and individualization in connectome development are related to psychiatric disorders. *Nature Neuroscience*, *20*(4), 513–515. https://doi.org/10.1038/nn.4511

Kotz, S. A. (2009). A critical review of ERP and fMRI evidence on L2 syntactic processing. *Brain and Language*, *109*(2–3), 68–74. https://doi.org/10.1016/j.bandl.2008.06.002

Li, P., Legault, J., & Litcofsky, K. A. (2014). Neuroplasticity as a function of second language learning: Anatomical changes in the human brain. *Cortex*, *58*, 301–324. https://doi.org/10.1016/j.cortex.2014.05.001

Liang, J., Wang, S.-J., & Zhou, C. (2021). Less is More: Wiring-Economical Modular Networks Support Self-Sustained Firing-Economical Neural Avalanches for Efficient Processing. *National Science Review*, nwab102. https://doi.org/10.1093/nsr/nwab102

Liu, C., de Bruin, A., Jiao, L., Li, Z., & Wang, R. (2021). Second language learning

tunes the language control network: A longitudinal fMRI study. *Language, Cognition and Neuroscience*, *36*(4), 462–473. https://doi.org/10.1080/23273798.2020.1856898

Liu, C., Jiao, L., Timmer, K., & Wang, R. (2021). Structural brain changes with second language learning: A longitudinal voxel-based morphometry study. *Brain and Language*, *222*, 105015. https://doi.org/10.1016/j.bandl.2021.105015

Liu, J., Liao, X., Xia, M., & He, Y. (2018). Chronnectome fingerprinting: Identifying individuals and predicting higher cognitive functions using dynamic brain connectivity patterns. *Human Brain Mapping*, *39*(2), 902–915. https://doi.org/10.1002/hbm.23890

Liu, L., Yan, X., Li, H., Gao, D., & Ding, G. (2020). Identifying a supramodal language network in human brain with individual fingerprint. *NeuroImage*, *220*, 117131. https://doi.org/10.1016/j.neuroimage.2020.117131

Liu, W., Wei, D., Chen, Q., Yang, W., Meng, J., Wu, G., Bi, T., Zhang, Q., Zuo, X.-N., & Qiu, J. (2017). Longitudinal test-retest neuroimaging data from healthy young adults in southwest China. *Scientific Data*, *4*(1), 170017. https://doi.org/10.1038/sdata.2017.17

Luk, G., Green, D. W., Abutalebi, J., & Grady, C. (2012). Cognitive control for language switching in bilinguals: A quantitative meta-analysis of functional neuroimaging studies. *Language and Cognitive Processes*, *27*(10), 1479–1488. https://doi.org/10.1080/01690965.2011.613209

Marshall, N., Timme, N. M., Bennett, N., Ripp, M., Lautzenhiser, E., & Beggs, J. M.

(2016). Analysis of Power Laws, Shape Collapses, and Neural Complexity: New Techniques and MATLAB Support via the NCC Toolbox. *Frontiers in Physiology*, *7*. https://doi.org/10.3389/fphys.2016.00250

Mei, L., Xue, G., Lu, Z.-L., Chen, C., Wei, M., He, Q., & Dong, Q. (2015). Long-term experience with Chinese language shapes the fusiform asymmetry of English reading. *NeuroImage*, *110*, 3–10. https://doi.org/10.1016/j.neuroimage.2015.01.030

Menéndez, M. L., Pardo, J. A., Pardo, L., & Pardo, M. C. (1997). The Jensen-Shannon divergence. *Journal of the Franklin Institute*, *334*(2), 307–318. https://doi.org/10.1016/S0016-0032(96)00063-4

Munsell, B. C., Gleichgerrcht, E., Hofesmann, E., Delgaizo, J., McDonald, C. R., Marebwa, B., Styner, M. A., Fridriksson, J., Rorden, C., Focke, N. K., Gilmore, J. H., & Bonilha, L. (2020). Personalized connectome fingerprints: Their importance in cognition from childhood to adult years. *NeuroImage*, *221*, 117122. https://doi.org/10.1016/j.neuroimage.2020.117122

Pliatsikas, C. (2020). Understanding structural plasticity in the bilingual brain: The Dynamic Restructuring Model. *Bilingualism: Language and Cognition*, *23*(2), 459–471. https://doi.org/10.1017/S1366728919000130

Połczyńska, M. M., & Bookheimer, S. Y. (2021). General principles governing the amount of neuroanatomical overlap between languages in bilinguals. *Neuroscience & Biobehavioral Reviews*, *130*, 1–14. https://doi.org/10.1016/j.neubiorev.2021.08.005

Ravindra, V., Drineas, P., & Grama, A. (2019). Constructing Compact Brain Connectomes for Individual Fingerprinting. *ArXiv:1805.08649 [Cs]*. http://arxiv.org/abs/1805.08649

Regalado, E. L., Schafer, W., McClain, R., & Welch, C. J. (2013). Chromatographic resolution of closely related species: Separation of warfarin and hydroxylated isomers. *Journal of Chromatography A*, *1314*, 266–275. https://doi.org/10.1016/j.chroma.2013.07.092

Spear, L. P. (2013). Adolescent Neurodevelopment. *Journal of Adolescent Health*, *52*(2), S7–S13. https://doi.org/10.1016/j.jadohealth.2012.05.006

Sporns, O., & Honey, C. J. (2006). Small worlds inside big brains. *Proceedings of the National Academy of Sciences*, *103*(51), 19219–19220. https://doi.org/10.1073/pnas.0609523103

Sun, X., Li, L., Ding, G., Wang, R., & Li, P. (2019). Effects of language proficiency on cognitive control: Evidence from resting-state functional connectivity. *Neuropsychologia*, *129*, 263–275. https://doi.org/10.1016/j.neuropsychologia.2019.03.020

Tang, H., Lu, X., Cui, Z., Feng, C., Lin, Q., Cui, X., Su, S., & Liu, C. (2018). Resting-state Functional Connectivity and Deception: Exploring Individualized Deceptive Propensity by Machine Learning. *Neuroscience*, *395*, 101–112. https://doi.org/10.1016/j.neuroscience.2018.10.036

Tononi, G., Sporns, O., & Edelman, G. M. (1994). A measure for brain complexity: Relating functional segregation and integration in the nervous system.

*Proceedings of the National Academy of Sciences*, *91*(11), 5033–5037. https://doi.org/10.1073/pnas.91.11.5033

van den Heuvel, M. P., & Sporns, O. (2011). Rich-Club Organization of the Human Connectome. *Journal of Neuroscience*, *31*(44), 15775–15786. https://doi.org/10.1523/JNEUROSCI.3539-11.2011

Wang, J., Wang, X., Xia, M., Liao, X., Evans, A., & He, Y. (2015). GRETNA: A graph theoretical network analysis toolbox for imaging connectomics. *Frontiers in Human Neuroscience*, *9*. https://doi.org/10.3389/fnhum.2015.00386

Wang, R., Fan, X., Liu, C., & Cai, Z. G. (2016). Cognitive control and word recognition speed influence the Stroop effect in bilinguals: BILINGUALS STROOP EFFECT. *International Journal of Psychology*, *51*(2), 93–101. https://doi.org/10.1002/ijop.12115

Wu, J., Yang, J., Chen, M., Li, S., Zhang, Z., Kang, C., Ding, G., & Guo, T. (2019). Brain network reconfiguration for language and domain-general cognitive control in bilinguals. *NeuroImage*, *199*, 454–465. https://doi.org/10.1016/j.neuroimage.2019.06.022

Xia, M., Wang, J., & He, Y. (2013). BrainNet Viewer: A Network Visualization Tool for Human Brain Connectomics. *PLoS ONE*, *8*(7), e68910. https://doi.org/10.1371/journal.pone.0068910

Yeh, F.-C., Vettel, J. M., Singh, A., Poczos, B., Grafton, S. T., Erickson, K. I., Tseng, W.-Y. I., & Verstynen, T. D. (2016). Quantifying Differences and Similarities in Whole-Brain White Matter Architecture Using Local Connectome Fingerprints.

*PLOS Computational Biology*, *12*(11), e1005203. https://doi.org/10.1371/journal.pcbi.1005203

Zhang, J., Zhuang, L., Jiang, J., Yang, M., Li, S., Tang, X., Ma, Y., Liu, L., & Ding, G. (2022). Brain fingerprints along the language hierarchy. *Frontiers in Human Neuroscience*, *16*, 982905. https://doi.org/10.3389/fnhum.2022.982905

Zuo, X.-N., & Xing, X.-X. (2014). Test-retest reliabilities of resting-state FMRI measurements in human brain functional connectomics: A systems neuroscience perspective. *Neuroscience & Biobehavioral Reviews*, *45*, 100–118. https://doi.org/10.1016/j.neubiorev.2014.05.009